\documentclass[aps,pra,twocolumn,showpacs]{revtex4-1}
\usepackage[OT4]{fontenc}

\usepackage{graphicx}
\usepackage{epstopdf}

\usepackage{amsmath,bm,amsfonts}

\begin{document}
\author{Piotr Kaczmarkiewicz$^{1,2}$}\email{piotr.kaczmarkiewicz@pwr.wroc.pl}
\author{Pawe{\l} Machnikowski$^{1}$}
\author{Tilmann Kuhn$^{2}$} 
\affiliation{$^{1}$Institute of Physics, Wroc{\l}aw University of Technology, 50-370 Wroc{\l}aw, Poland}
\affiliation{$^{2}$Institut f\"ur Festk\"orpertheorie, Westf\"alische Wilhelms-Universit\"at, Wilhelm-Klemm-Str.~10, 48149~M\"unster, Germany }
\title{Carrier trapping in a quantum dash: optical signatures}

\begin{abstract}
We theoretically study the optical properties and the electronic structure of highly elongated quantum dots (quantum dashes) and show how geometrical fluctuations affect the excitonic spectrum of the system. The dependence of the absorption intensities on the geometrical properties (depth and length) of the trapping center in a quantum dash is analyzed and the dependence of the degree of the linear polarization on these geometrical parameters is studied.
\end{abstract}
\maketitle
\section{Introduction}

Quantum dashes (QDashes) are highly elongated self-assembled quantum dot structures  \cite{reithmaier07}. They are characterized by high surface density, emission wavelength in the region of the third telecommunication window, broad gain and tunability  \cite{reithmaier07,dery04,sauerwald05,hein09}. They are particularly interesting from the point of telecommunication applications, where they are now commonly used (in InP material systems) to realize high quality lasers and optical amplifiers operating at $1.55~\mathrm{{\mu}m}$  \cite{reithmaier07,lelarge07}. 

Structural data  \cite{reithmaier07} reveal that the geometrical shape of the real QDashes is non-uniform, with zig-zag bends and cross-section size fluctuations in the form of local widenings. Both these shape irregularities, as well as compositional inhomogeneities and the related non-uniform strain distribution, can induce an additional trapping potential within a QDash structure. Observed polarization properties of a QDash ensemble can be explained only if such an additional carrier trapping is assumed within the QDash volume  \cite{musial12}. 

Both electronic properties as well as optical transitions of uniformly-shaped \cite{miska04,planelles09,andrzejewski10,saito08,sheng08} QDashes have been modeled previously. Also optical properties of non-uniformly shaped \cite{musial12,kaczmarkiewicz12} quantum dashes have been modeled, however only some of the geometrical factors of the QDash trapping center have been addressed in those studies. 
-
In this paper, we extend our previous work to the case where not only the depth of the trapping potential is varied but also its length. We study how both those parameters characterizing the local widening of a QDash influence the optical properties of an exciton localized in the system. Additional trapping inside the elongated structure of a QDash strongly affects the optical properties of the system, hence, we find that in order to fully characterize carrier trapping in a QDash systems all geometrical factors have to be studied. 

The paper is organized as follows. In Sec. \ref{s:model}, we provide the model and simplified theoretical framework of the studied system. In Sec. \ref{s:results}, we present the results of our theoretical analysis concerning the electronic structure and optical properties of the system. We conclude the paper in Sec. \ref{s:conclusions}.

\section{The Model}\label{s:model}

In our study, we consider a highly elongated quantum dot-like structure with the length to width ratio greater than 5 and characterized by the width and thickness variation along its length, located symmetrically in the center of the structure (preserving $D_2$ symmetry). Such a thickness variation provides an additional confinement within the structure and has been successful in qualitatively reproducing the temperature dependence of the degree of linear polarization (DOP) of an ensemble of InAs/InP quantum dashes \cite{musial12}.
\begin{figure}[tb]
\begin{center}
\includegraphics[width=70mm]{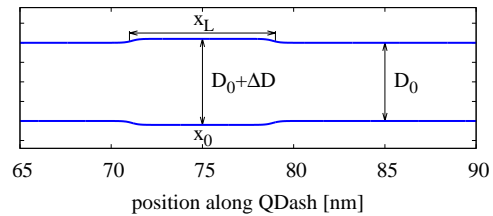}
\end{center}
\caption{\label{fig:schemat}(Color online) Top view of the region of the structure widening. The cross section is circular segment, with fixed height to chord base length ratio of 1:5.5. The total length of the structure is $L=150$ nm.}
\end{figure}

In our modeling, we use the single-band effective mass approximation and envelope wave function description. The Hamiltonian of a single carrier (electron or hole) is then
\begin{equation}\label{eq:ham}
H=-\frac{\hbar^2}{2m^*} \Delta + V(\bm{r}),
\end{equation}
where $m^*$ is the effective mass of the carrier and the characteristic features of the structure geometrical properties are described by the potential term $V(\bm{r})$. 

The cross section of the structure is assumed to be a circular segment, with the base width changing along the QDash length according to the model function (see Fig.~\ref{fig:schemat} for visualization)
\begin{displaymath}
D(x)= D_{0}+\frac{\Delta D (1+4e^{-b})}{1+4e^{-b}\cosh[2b(x-x_0)/x_{L}]},
\end{displaymath}
where $D_0$ is the QDash base width away from the widening, $\Delta D$ is the depth of the fluctuation, $x_L$ is the length of the fluctuation and $x_0$ is the position of the center of the widening. The widening parameter $\lambda=\Delta D/D_0$ is defined as the ratio of the excess width to the QDash width away from the trapping center. The QDash width to height ratio is kept constant, $D = \alpha H$, with $\alpha=5.5$, which is typical for these structures \cite{sauerwald05}. The total length of the structure is set to $150$ nm, and the length to width ratio is $6:1$. The material parameters used in our calculations are those for InAs/InP structures  \cite{lawaetz71,musial12}. 

In order to find single carrier envelope wave functions we follow the adiabatic approximation \cite{wojs96} and separate the directions of the weakest ($x$) confinement from the other directions ($y$, $z$). Then, we use the variational principle and minimize the Hamiltonian (\ref{eq:ham}) in the basis of two dimensional harmonic oscillator ($\psi_{n}(x;y,z)$) on a fixed grid along the QDash elongation direction ($x$). 
In this way, we obtain approximate energies $\epsilon_{n}(x) = \langle \psi_{n}(x;y,z) | H_{c}|\psi_{n}(x;y,z)\rangle$ which are treated as an effective potential for the one-dimensional eigenvalue equation in the QDash elongation direction,
\begin{equation}\label{wzor:RnieEfektywn}
\left [ -\frac{\hbar^{2}}{2m^*}\frac{\partial^{2}}{\partial x^{2}} 
+ \epsilon(x) \right ] \phi_{nm}(x) = E_{nm} \phi_{nm}(x),
\end{equation}
which is then solved numerically. 

Based on the calculated single-carrier states, one can construct the product basis for the excitonic states 
and diagonalize the system described by the Hamiltonian
\begin{equation}
H = \sum_{i} E_{i}^{(e)} a_{i}^\dag a_{i}+\sum_{i}
E_{i}^{(h)} h_{i}^\dag h_{i } + \sum_{ijkl} V_{ijkl} a_{i}^\dag h_{j}^\dag h_{k} a_{l}, \nonumber
\label{wzor:ham_ex}
\end{equation}
where 
$a_{i}^{\dag}a_{i}$ and $h_{i}^{\dag}h_{i}$ are electron and hole
creation and annihilation operators, respectively,
$E_{i}^{e,h}$ are the energies found from
Eq.~\eqref{wzor:RnieEfektywn} for electrons and holes,  
and $V_{ijkl}$ are the matrix elements of the electron-hole
interaction.

In calculations of the dipole moments, mainly heavy hole character of the hole states is assumed, with only a small admixture of light holes (see Ref. \onlinecite{musial12} for details). The interband dipole moments for the direction parallel ($l$) and transverse ($t$) to the direction of the elongation of the structure are 
\begin{equation}\label{wzor:dl}
d_{l,t}^{(\beta)} = \mp d_0 \frac{i\pm1}{2}\alpha_{3/2,1/2}^{(\beta)}
+ d_0 \frac{1\mp i}{2\sqrt{3}}\alpha_{-1/2,1/2}^{(\beta)}\nonumber,
\end{equation}
where the upper and lower signs correspond to '$l$' and '$t$' respectively and the parameters $\alpha_{-1/2,1/2}$ and $\alpha_{3/2,1/2}$ are the oscillator strengths for light and heavy hole contributions.

The geometry of the structure is reflected in its optical properties by the degree of linear polarization, which for a given state $\beta$ is 
\begin{equation}
P^{\beta}= (|d_{l}^{\beta}|^{2}-|d_{t}^{\beta}|^{2})/(|d_{l}^{\beta}|^{2}+|d_{t}^{\beta}|^{2}). \nonumber
\end{equation}

\section{Results}\label{s:results}
In this section, we present the results of the theoretical modeling of a single exciton confined in a QDash structure with an additional trapping center present. We change the properties of the trapping center by tuning two geometrical parameters, the widening parameter $\lambda$, correlated with the depth of the trapping center, and widening length $x_{L}$ defining the length of the widened sector of a QDash.

\begin{figure}[tb]
\begin{center}
\includegraphics[width=70mm]{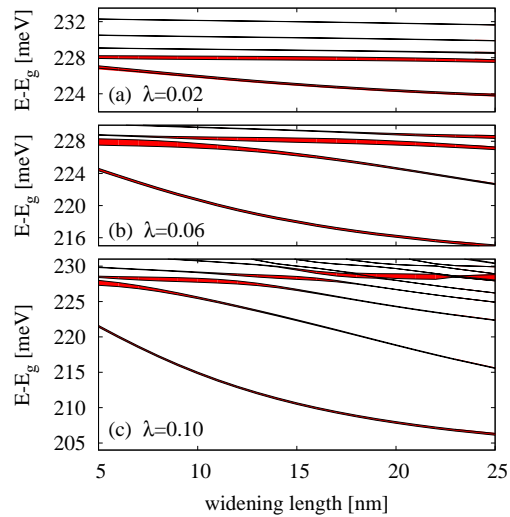}
\end{center}
\caption{\label{fig:spectr}(Color online) Exciton energy spectra as a function of the QDash widening length $x_{L}$ for three different sizes of width fluctuation $\lambda$. The linewidth is proportional to the total absorption intensity $\sim |d_{l}|^{2}+|d_{t}|^{2}$. }
\end{figure}

\begin{figure}[tb]
\begin{center}
\includegraphics[width=70mm]{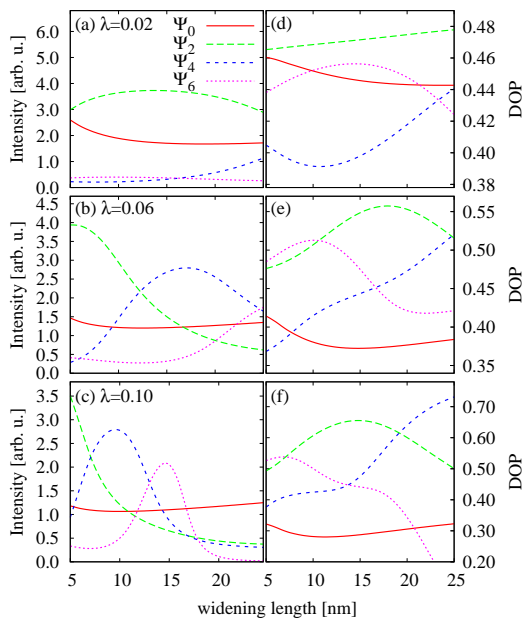}
\end{center}
\caption{\label{fig:dop}(Color online) Absorption intensity [(a)-(c)] and the degree of linear polarization [(d)-(f)] of the four lowest bright exciton states for tree different values of the widening parameter $\lambda$ as a function of the length of the widening $x_{L}$. The dark states constructed with electron and hole wave functions of opposite parity (odd indices) have not been shown.}
\end{figure}

In Fig. \ref{fig:spectr}, we compare the energy spectra for three different values of $\lambda$ as a function of the widening length.
As can be seen in Fig. \ref{fig:spectr} (a), the energy shifts for the presented range of widening lengths are relatively small. For the two extreme widening lengths presented in the figure, the total shift of the ground state is only $3~\mathrm{meV}$. It is understandable, as in this case the trapping center is relatively shallow and increasing its length has only a minor effect on the system spectrum. The energy shift grows larger when we increase the trapping center depth. In Fig. \ref{fig:spectr} (b) ($\lambda=0.06$), the ground state energy decreases by $10~\mathrm{meV}$ and in Fig. \ref{fig:spectr} (c) ($\lambda=0.10$) by $15~\mathrm{meV}$. 
The energies of several lowest lying excitonic states also decrease strongly but their energy shifts are smaller than that of the ground state. Apart from understandable down-shifts of exciton energies one can also observe non-trivial changes in absorption intensities. In Figs. \ref{fig:dop} (a-c), we present the absorption intensities for four lowest energy bright excitonic states. Surprisingly, even for large values of the widening parameter (Fig.~\ref{fig:dop} (c)) there is little to no change in the intensity of the ground state. The reason is twofold: the energy separation between the ground state and higher excitonic states is relatively large and therefore the ground state consists mostly of the lowest energy exciton basis state with only small admixture of higher states and the overlap between electron and hole in the ground state changes only slightly in the presented range of widening lengths.

One can also see a strong decline (Fig. \ref{fig:dop} (b-c)) in the intensity of the first excited bright state ($\Psi_{2}$). Again, two factors play a role here: on one hand the overlap between the electron and the hole changes with the change in the length of the trapping center but, more importantly, the Coulomb interaction leads to a change in the composition  of this excitonic state. Similar, non-trivial behavior can be observed for higher excitonic states when a large enhancement of the absorption intensity is visible followed by a strong decline. 

In  Fig. \ref{fig:spectr} (d-f), we show the polarization properties of exciton eigenstates. Only a small change in the DOP of the exciton ground state is observed, even though the widening length should be a major factor in the polarization properties of this state, as the degree of linear polarization can be estimated with a simplified model  \cite{kaczmarkiewicz11a} as $P \sim (1/D^{2} -1/L^{2})$. The reason for such a small change in the DOP of the ground state is that in most of the presented region of widening lengths the carrier wave function spans over a much larger area than the widening length itself. It is especially visible for very small values of $x_{L}$ where the confinement of the carriers is even weaker and hence the DOP is larger. When the widening length increases to larger values the DOP also increases as was predicted by the aforementioned model.
The width of the trapping center has a very strong influence on the polarization properties of higher energy excitonic states. The degree of polarization of the second excited bright state ($\Psi_{4}$) increases almost by a factor of two (for $\lambda=0.10$) between moderate (5--10~nm) and large (25~nm) values of the widening length. The third excited state ($\Psi_{6}$) changes non-monotonically and decreases strongly for $\lambda=0.10$, although for widening lengths larger than 20~nm it has nearly zero intensity.

\section{Conclusions}\label{s:conclusions}
We have investigated the influence of the trapping center 
length and its depth on the electronic structure and optical properties 
of an inhomogeneously shaped single quantum dash. We have shown that changing the  length of the trapping
center leads to non-trivial changes in the QDash exciton spectrum.
The change in the widening length leads not only to expected down-shift
in exciton energy but also to large changes in the absorption intensities.
On the other hand, both the exciton ground state intensity and its 
degree of polarization changes only slightly when the widening length is varied.
The scale of the observed changes in the optical properties is larger when the trapping
potential is deeper.  
We have also shown that individual properties of QDashes highly depend on the 
details of the confining potential and both the trapping center length and depth
are important when modeling such structures. 

\begin{acknowledgments}
We acknowledge support from the TEAM programme of the 
Foundation for Polish Science, co-financed by the European Regional
Development Fund, and from the Alexander von Humboldt Foundation.
PK acknowledges support from the German Academic Exchange Service (DAAD).
\end{acknowledgments}

\bibliographystyle{prsty}
\bibliography{abbr,quantum}

\begin{thebibliography}{10}

\bibitem{reithmaier07}
J.~P. Reithmaier, G. Eisenstein, and A. Forchel, Proc. IEEE {\bf 95},  1779
  (2007).

\bibitem{dery04}
H. Dery, E. Benisty, A. Epstein, R. Alizon, V. Mikhelashvili, G. Eisenstein, R.
  Schwertberger, D. Gold, J.~P. Reithmaier, and A. Forchel, J. Appl. Phys. {\bf
  95},  6103  (2004).

\bibitem{sauerwald05}
A. Sauerwald, T. Kummell, G. Bacher, A. Somers, R. Schwertberger, J.~P.
  Reithmaier, and A. Forchel, Appl. Phys. Lett. {\bf 86},  253112  (2005).

\bibitem{hein09}
S. Hein, S. Hofling, and A. Forchel, IEEE Phot. Tech. Lett. {\bf 21},  528
  (2009).

\bibitem{lelarge07}
F. Lelarge, B. Dagens, J. Renaudier, R. Brenot, A. Accard, F. van Dijk, D.
  Make, O.~L. Gouezigou, J.-G. Provost, F. Poingt, J. Landreau, O. Drisse, E.
  Derouin, B. Rousseau, F. Pommereau, and G.-H. Duan, IEEE J. Sel. Top. Quant.
  El. {\bf 13},  111   (2007).

\bibitem{musial12}
A. Musia\l{}, P. Kaczmarkiewicz, G. S\ifmmode~\mbox{\k{e}}\else \k{e}\fi{}k, P.
  Podemski, P. Machnikowski, J. Misiewicz, S. Hein, S. H\"ofling, and A.
  Forchel, Phys. Rev. B {\bf 85},  035314  (2012).

\bibitem{miska04}
P. Miska, J. Even, C. Platz, B. Salem, T. Benyattou, C. Bru-Chevalier, G.
  Guillot, G. Bremond, K. Moumanis, F.~H. Julien, O. Marty, C. Monat, and M.
  Gendry, J. Appl. Phys. {\bf 95},  1074  (2004).

\bibitem{planelles09}
J. Planelles, M. Royo, A. Ballester, and M. Pi, Phys. Rev. B {\bf 80},  045324
  (2009).

\bibitem{andrzejewski10}
J. Andrzejewski, G. S{\k e}k, E. O'Reilly, A. Fiore, and J. Misiewicz, J. Appl.
  Phys. {\bf 107},  073509  (2010).

\bibitem{saito08}
T. Saito, H. Ebe, Y. Arakawa, T. Kakitsuka, and M. Sugawara, Phys. Rev. B {\bf
  77},  195318  (2008).

\bibitem{sheng08}
W. Sheng and S.~J. Xu, Phys. Rev. B {\bf 77},  113305  (2008).

\bibitem{kaczmarkiewicz12}
P. Kaczmarkiewicz and P. Machnikowski, arXiv:1203.3977 [cond-mat.mes-hall]
  (unpublished).

\bibitem{lawaetz71}
P. Lawaetz, Phys. Rev. B {\bf 4},  3460  (1971).

\bibitem{wojs96}
A. W\'ojs, P. Hawrylak, S. Fafard, and L. Jacak, Phys. Rev. B {\bf 54},  5604
  (1996).

\bibitem{kaczmarkiewicz11a}
P. Kaczmarkiewicz, A. Musia{\l}, G. S{\k{e}}k, P. Podemski, P. Machnikowski,
  and J. Misiewicz, Acta Phys. Pol. A {\bf 119},  633  (2011).

\end{thebibliography}
\end{document}